\documentclass[aps,pra,twocolumn,superscriptaddress]{revtex4-1}

\usepackage{graphicx}
\usepackage{hyperref}
\usepackage{amsmath}
\usepackage{amssymb}
\usepackage{epstopdf}
\usepackage{multirow}
\usepackage{threeparttable}
\usepackage[usenames]{color}

\UseRawInputEncoding
\begin{document}

\title{Single-photon time-stretch computational ghost spectroscopy}

\author{Zhibin Zhao}
\affiliation{State Key Laboratory of Precision Spectroscopy, and Hainan Institute, East China Normal University, Shanghai 200062, China}

\author{Kun Huang}
\email{khuang@lps.ecnu.edu.cn}
\affiliation{State Key Laboratory of Precision Spectroscopy, and Hainan Institute, East China Normal University, Shanghai 200062, China}
\affiliation{Chongqing Key Laboratory of Precision Optics, Chongqing Institute of East China Normal University, Chongqing 401121, China}
\affiliation{Collaborative Innovation Center of Extreme Optics, Shanxi University, Taiyuan, Shanxi 030006, China}

\author{Ben Sun}
\affiliation{State Key Laboratory of Precision Spectroscopy, and Hainan Institute, East China Normal University, Shanghai 200062, China}

\author{Beibei Dong}
\affiliation{State Key Laboratory of Precision Spectroscopy, and Hainan Institute, East China Normal University, Shanghai 200062, China}

\author{Wen Zhang}
\affiliation{State Key Laboratory of Precision Spectroscopy, and Hainan Institute, East China Normal University, Shanghai 200062, China}

\author{Jianan Fang}
\affiliation{State Key Laboratory of Precision Spectroscopy, and Hainan Institute, East China Normal University, Shanghai 200062, China}
\affiliation{Chongqing Key Laboratory of Precision Optics, Chongqing Institute of East China Normal University, Chongqing 401121, China}

\author{Heping Zeng}
\email{hpzeng@phy.ecnu.edu.cn}
\affiliation{State Key Laboratory of Precision Spectroscopy, and Hainan Institute, East China Normal University, Shanghai 200062, China}
\affiliation{Chongqing Key Laboratory of Precision Optics, Chongqing Institute of East China Normal University, Chongqing 401121, China}
\affiliation{Shanghai Research Center for Quantum Sciences, Shanghai 201315, China}
\affiliation{Chongqing Institute for Brain and Intelligence, Guangyang Bay Laboratory, Chongqing, 400064, China}

\begin{abstract}
Time-stretch spectroscopy is powerful for capturing transient spectral phenomena but remains fundamentally limited by detector bandwidth or timing jitter, especially under photon-starved conditions. Here, we devise and implement single-photon time-stretch computational ghost spectroscopy, which integrates dispersive wavelength-to-time mapping with programmable temporal encoding and correlation-based reconstruction to overcome these detection limitations. Specifically, temporally stretched ultrashort pulses are modulated by predefined encoding patterns and detected by a low-bandwidth detector, allowing reconstruction of near-infrared spectra with 450 resolvable channels across 1530-1590 nm without direct high-speed waveform acquisition. By further incorporating compressive sensing, accurate spectral recovery is achieved at sub-Nyquist sampling rates, substantially reducing acquisition requirements to facilitate high-speed operation at 210 kHz. In the single-photon regime, computational ghost reconstruction effectively suppresses the intrinsic detector timing jitter, yielding high-fidelity spectra at illumination fluxes down to 0.01 photons/pulse. By jointly enabling broadband coverage, high spectral resolution, high acquisition speed, and single-photon sensitivity, this approach establishes a computation-enhanced paradigm for time-stretch spectroscopy and provides a versatile platform for ultrafast and photon-efficient spectroscopic applications.

\end{abstract}

\maketitle

\section{Introduction}
Optical spectroscopy is a versatile analytical tool to provide critical information on sample composition, structure, and chemical interactions \cite{Ozaki2021Book}. Owing to its rapid, non-destructive, and non-invasive nature, it has found widespread applications in chemical analysis, biomedical diagnostics, and agricultural quality control \cite{Hakkel2022NC, Afara2021NP, Tsuchikawa2022AS}. Conventional grating-based spectrometers spatially disperse spectra onto detector arrays, but their acquisition speed is fundamentally limited by detector readout rates and an inherent trade-off between spectral resolution and bandwidth imposed by the finite number of pixels \cite{Yang2021Science, Zhang2025eLight, Zhu2022JNIS}. Fourier-transform infrared (FTIR) spectroscopy offers broadband and high-resolution measurements \cite{Griffiths1983Science}, yet its speed is constrained by mechanical delay scanning, typically limiting frame rates to the kilohertz regime even in advanced implementations \cite{Hashimoto2018NC}. Dual-comb spectroscopy enables scan-free and high-resolution acquisition \cite{Xu2024Nature, Long2024NP}, but its practical deployment generally relies on complex and tightly stabilized laser systems. In addition, Fourier-domain mode-locked (FDML) lasers provide a powerful platform for high-speed spectroscopy \cite{Kranendonk2007OptExpress, Huang2022Sensors, Cai2026arXiv, Jiang2020NP}, although the achievable spectral bandwidth is typically constrained by the gain bandwidth of the wavelength-swept laser source. Consequently, realizing infrared spectroscopy that achieves high speed, broad bandwidth, and high resolution in a compact and versatile architecture remains a long-standing challenge.

Photonic time-stretch spectroscopy, also known as dispersive Fourier transform (DFT) \cite{Solli2008NP}, provides an elegant solution by mapping spectral information into the time domain, enabling single-shot spectral acquisition with a single-pixel detector \cite{Goda2013NP, Mahjoubfar2017NP, Zhang2025NRMP}. This approach has proven particularly powerful for capturing non-repetitive and transient events \cite{Kawai2020CommPhys, Lei2018NP, Zhou2025Light}, and has recently been extended to the single-photon regime via time-correlated single-photon counting (TCSPC) \cite{Cai2024SA, Yang2022SB, Sun2024LPR, Avenhaus2009OL}. Despite these advances, conventional time-stretch spectroscopy remains fundamentally constrained by detector bandwidth and timing jitter \cite{Fard2013LPR, Sun2024LPR, Tiedeck2022ACSPhotonics}. Accurate reconstruction of rapidly varying stretched waveforms requires ultrafast photodetectors and high-bandwidth digitizers \cite{Hashimoto2023LSA, He2024APL}, which limit the number of resolvable spectral channels and the achievable resolution after wavelength-to-time mapping. These constraints are especially severe in the single-photon regime, where the intrinsic timing jitter of infrared photon-counting detectors markedly degrades spectral resolution \cite{Sun2024LPR}. 

In parallel, computational spectroscopy has emerged as a powerful paradigm for enhancing spectral performance while relaxing hardware complexity through algorithmic reconstruction \cite{Redding2013NP, Chen2024LSA, Kong2021NL, Cheng2019NC, Zhang2024NC, Xiao2022ACSP, Burghoff2016SciAdv}. In particular, temporal computational ghost imaging exploits structured temporal modulation and correlation-based reconstruction to recover fast signals using slow detectors, effectively shifting the bandwidth requirement from the detector to the modulator and computation \cite{Ryczkowski2016NP, Devaux2016Optica, Zhang2025LPR}. Building on this, recent studies have applied temporal ghost imaging techniques to spectroscopic measurements \cite{Amiot2018OL, Peng2025NC, Hu2025OL, Rabi2021IEEE, Sanna2024AQT, Janassek2018PRA, Zhao2024JLT}, achieving high-speed, high-resolution spectral acquisition and enhanced measurement dimensionality. Inspired by these studies, the strong conceptual complementarity between time-stretch spectroscopy and computational temporal modulation enables substantial relaxation of detector bandwidth and timing jitter constraints, with particularly pronounced advantages in the single-photon regime. This strategy thus provides a viable pathway toward realizing high spectral resolution, broad bandwidth, and rapid acquisition without the need for ultrafast and low timing jitter detectors.

 In this work, we demonstrate high-resolution and high-sensitivity time-stretch computational ghost spectroscopy by integrating dispersive wavelength-to-time mapping with programmable temporal encoding and correlation-based reconstruction. Unlike conventional time-stretch spectroscopy, which relies on direct temporal waveform sampling and is fundamentally constrained by detector bandwidth and timing jitter, the proposed approach enables high-fidelity spectral recovery through structured temporal encoding and computation-assisted reconstruction. More fundamentally, it transforms spectral acquisition from localized time-resolved detection into a multiplexed encoding-and-decoding process, in which spectral information is acquired through patterned temporal projections rather than direct waveform measurement. In our implementation, temporally stretched ultrashort pulses are encoded using predefined temporal patterns and detected with a low-bandwidth detector, enabling reconstruction of NIR spectra with 450 resolvable channels across 1530-1590 nm. By incorporating compressive sensing and warped encoding, the acquisition burden is substantially reduced while maintaining high spectral resolution, enabling high-speed operation at 210 kHz. In the single-photon regime, computational ghost reconstruction effectively mitigates the intrinsic timing jitter of NIR photon-counting detectors to yield high-fidelity spectra at incident illumination fluxes down to 0.01 photons/pulse. This work establishes a photon-efficient, computation-enhanced paradigm for time-stretch spectroscopy that enables high-resolution spectral measurements under photon-limited conditions.

\begin{figure*}[t!]
\centering
\includegraphics[width=0.92\textwidth]{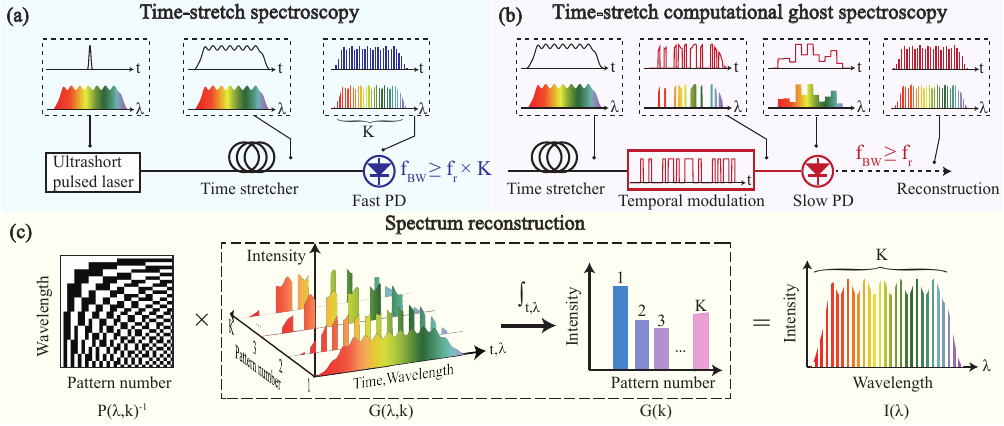}
\caption{Conceptual comparison of conventional and computational time-stretch spectroscopy. (a) Conventional time-stretch spectroscopy. A broadband optical pulse is dispersed in a fiber, mapping its spectrum into the time domain. Accurate spectral reconstruction relies on direct detection of the rapidly varying temporal waveform and is therefore fundamentally limited by the electronic bandwidth of the detector. (b) Proposed time-stretch computational ghost spectroscopy. The time-stretched pulse is modulated by a sequence of predefined temporal encoding patterns. A low-bandwidth photodetector records only the pattern-dependent integrated intensity, rather than the full temporal waveform, thereby circumventing the detector bandwidth limitation. (c) Ghost reconstruction of the spectrum. The input spectrum is computationally retrieved through correlation between the measured integrated intensities and the known temporal encoding patterns.}
\label{fig1}
\end{figure*}

\section{Basic principle}
Figure \ref{fig1} compares conventional time-stretch spectroscopy and the proposed time-stretch computational ghost spectroscopy. In conventional time-stretch spectroscopy, the spectrum of an ultrashort optical pulse is mapped into the time domain through dispersive propagation, enabling single-shot spectral measurements via temporal detection \cite{Mahjoubfar2017NP, Hashimoto2023LSA, Kawai2020CommPhys}, as shown in Fig. \ref{fig1}(a). Under the assumption of dominant second-order dispersion, the temporal intensity profile at the output of the dispersive medium can be expressed as \cite{Goda2013NP}

\begin{equation}
|E(L, T)|^2 = \frac{2}{\pi \beta_2 L}  e^{-\alpha L}
\left| \tilde{E} \left(0, \frac{T}{\beta_2 L}\right) \right|^2 ,
\label{eq1}
\end{equation}

where $E$ denotes the field amplitude of the optical pulse, $\tilde{E}$ denotes its Fourier transform, $\alpha$ is the absorption coefficient, $L$ is the propagation length, $\beta_2$ is the second-order dispersion coefficient, and $T$ is the relative time in the moving frame of the propagation pulse. Equation (\ref{eq1}) establishes a linear wavelength-to-time mapping, such that a spectral interval $\delta\lambda$ is stretched into a temporal interval $\delta t = D L \,\delta\lambda$, where $D$ is the group velocity dispersion (GVD) per unit length introduced by the dispersive medium. This one-to-one mapping forms the physical basis of time-stretch spectroscopy.

Despite its capability for ultrafast single-shot measurements, conventional time-stretch spectroscopy imposes stringent requirements on the detection bandwidth. To faithfully resolve K spectral channels at a laser repetition rate $f_r$, the detector bandwidth $f_{\mathrm{BW}}$ must satisfy $f_{\mathrm{BW}} \ge f_r K$ \cite{Goda2013NP}. This requirement severely restricts spectral resolution when only low-bandwidth detectors are available. In the single-photon regime, the situation is further exacerbated by the intrinsic timing jitter of single-photon detectors, which directly degrades the achievable spectral fidelity \cite{Sun2024LPR}.

To overcome these limitations, we introduce time-stretch computational ghost spectroscopy [Fig. \ref{fig1}(b)], which combines dispersive spectral mapping with structured temporal intensity encoding. After the initial wavelength-to-time mapping, the stretched waveform is modulated by a sequence of predefined temporal patterns. Instead of directly sampling the rapidly varying waveform, a slow detector records only the integrated intensity associated with each modulation pattern. This strategy shifts the bandwidth requirement from the detector to the modulator and computational reconstruction, such that the detector bandwidth needs only to exceed the laser repetition rate $f_r$.

Denoting the stretched temporal waveform as $I(t)$ and exploiting the calibrated wavelength-to-time mapping, the spectral profile $I(\lambda)$ can be inferred from the pattern-dependent integrated measurements. The measurement process can be written in a discrete form as \cite{Gibson2020OE}
\begin{equation}
\mathbf{G} = \mathbf{P} \mathbf{I} \ ,
\label{eq2}
\end{equation}
where $\mathbf{G}$ is the vector of measured integrated intensities, $\mathbf{P}$ is the measurement (encoding) matrix determined by the temporal modulation patterns, and $\mathbf{I}$ represents the discretized spectrum to be reconstructed. Spectral retrieval is achieved by solving the corresponding inverse problem through correlation-based computational reconstruction, a hallmark of ghost spectroscopy \cite{Amiot2018OL}. This encoding-and-reconstruction framework enables high-resolution spectral recovery using low-bandwidth detectors and is intrinsically robust against detector timing jitter \cite{Peng2025NC, Hu2025OL, Rabi2021IEEE}.

Notably, the proposed framework naturally supports compressive sensing, allowing spectral reconstruction from a reduced number of measurements. When the spectrum is sparse or compressible in a suitable basis $\mathbf{\Psi}$, high-fidelity recovery can be achieved at sub-Nyquist sampling rates \cite{Duarte2008ICASSP}. The compressed measurement model is expressed as
\begin{equation}
\mathbf{G} = \mathbf{P}\mathbf{I}
= \mathbf{P}\mathbf{\Psi}\mathbf{s}
= \mathbf{\Theta}\mathbf{s},
\quad \mathbf{\Theta} = \mathbf{P}\mathbf{\Psi} \ ,
\label{eq3}
\end{equation}
where $\mathbf{s}$ denotes the sparse coefficient vector of the spectrum in the basis $\mathbf{\Psi}$. In our implementation, a Walsh-ordered Hadamard matrix is employed as the measurement matrix to ensure low coherence with the sparse basis, and spectral reconstruction is performed using the total-variation augmented Lagrangian (TVAL3) algorithm. This compressive strategy relaxes acquisition demands without compromising spectral resolution, enabling high-speed spectroscopic operation \cite{Vaz2020OE}. Further details of the reconstruction procedures are provided in Supplementary Note 1.

Notably, temporal modulation combined with computational reconstruction has previously been explored in photonic time-stretch imaging to enhance information recovery while relaxing detector bandwidth requirements \cite{Bosworth2015OE, Guo2015OE, Mididoddi2017IPJ, Chi2019IPJ, Li2023ACSP}. In contrast to these imaging-oriented implementations, the present work extends this computation-assisted acquisition framework to spectroscopy in detector-limited and photon-starved regimes, where detector timing jitter directly limits the achievable spectral resolution after wavelength-to-time mapping.

\begin{figure*}[t!]
	\centering
	\includegraphics[width=0.8\textwidth]{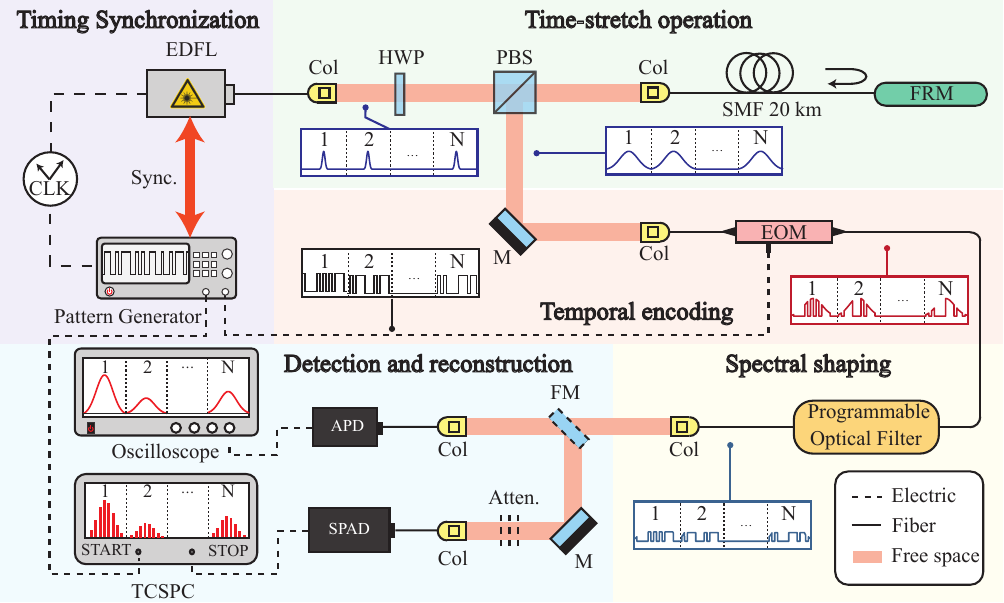}
	\caption{Experimental setup. Near-infrared pulses from a Er-doped fiber laser (EDFL) are polarization-conditioned by a half-wave plate (HWP) and a polarizing beam splitter (PBS), and subsequently temporally stretched in a 20 km single-mode fiber (SMF). A Faraday mirror reflects the pulses back through the same fiber in a double-pass configuration, mitigating polarization-mode dispersion and routing the signal via the PBS to a 20 GHz electro-optic modulator (EOM). The EOM, driven by a pulse pattern generator synchronized with the laser to a common 10 MHz reference clock, imposes programmable temporal intensity encodings on the stretched pulses. The encoded signal is then spectrally shaped by a programmable optical filter. Under strong-light conditions, the output is detected by a low-bandwidth avalanche photodiode (APD) and digitized by a real-time oscilloscope, whereas under weak-light conditions it is measured by a single-photon avalanche diode (SPAD) and processed using a time-correlated single-photon counting (TCSPC) module. M: mirror; DM: dichroic mirror; HWP: half-wave plate; Col: fiber collimator; FM: flipping mirror; FRM: Faraday mirror.}
	\label{fig2}
\end{figure*}

\section{Experimental Setup}
Figure \ref{fig2} illustrates the experimental setup of the time-stretch computational spectrometer. An erbium-doped mode-locked fiber laser (EDFL, LangyanTech, ErPico RLocking) operating at 1.56 $\mu$m served as the optical source. The laser repetition rate was tunable around 21.53 MHz through an intracavity piezo-actuator mounted on a motorized translation stage. By locking the laser to an external 10 MHz rubidium reference clock, the repetition rate was stabilized at 21.525 MHz and synchronized with the electrical pattern generator. 

The output pulses were polarization-conditioned and launched into a 20 km single-mode fiber (SMF, YOFC, G652D) for temporal stretching via a polarization beam splitter (PBS). The SMF exhibited an attenuation of approximately 0.22 dB km$^{-1}$ and a GVD about 17.1 ps nm$^{-1}$ km$^{-1}$. A Faraday mirror was placed at the fiber end reflects the pulses back through the same SMF in a double-pass configuration, automatically compensating fiber-induced polarization fluctuations and polarization-mode dispersion, while routing the returned signal to the PBS. This configuration also ensures that the polarization state of the reflected pulses is properly aligned with the principal axis of the polarization-maintaining electro-optic modulator (EOM), enabling stable and efficient high-speed temporal modulation. More experimental details on the electro-optic modulation can be found in Supplementary Note 2. After the round trip, the average optical power was approximately 1 mW. The resulting temporally stretched pulses provide a one-to-one wavelength-to-time mapping and form the basis for subsequent temporal encoding.

The stretched NIR pulses were modulated by a 20 GHz EOM (Keyang, KY-AM-15-20G), driven by a synchronized pulse pattern generator (Anritsu, MP1763C). With an electrical bandwidth of 12.5 GHz, the generator enabled programmable binary temporal encoding with a temporal resolution of 80 ps.

Following temporal modulation, the optical signal was spectrally shaped using a programmable optical filter (Finisar, Waveshaper 1000A) based on liquid-crystal-on-silicon technology. The filter provided independent control over spectral amplitude with programmable bandwidths ranging from 10 GHz to 5.36 THz and attenuation up to 35 dB. This capability allowed the generation of arbitrary spectral profiles, facilitating controlled emulation of diverse spectral signatures for system calibration and validation.

For detection under strong-light conditions, the encoded time-stretch signals were measured using a fiber-coupled InGaAs avalanche photodiode (APD, Keyang, KY-APRM-100M-I-FA) with a bandwidth of 100 MHz and digitized by a real-time oscilloscope (Keysight DSA91304A)  operating at 40 GS s$^{-1}$ with a 10 GHz analog bandwidth. Under photon-starved conditions, the signals were detected by a SPAD (LangyanTech, SPD-Infra) and recorded using a TCSPC (qutools, quTAG). In both cases, the spectral information was reconstructed through correlation-based computational processing between the measured pattern-dependent integrated intensities and the corresponding temporal encoding patterns.

\begin{figure*}[t!]
\centering
\includegraphics[width=0.75\textwidth]{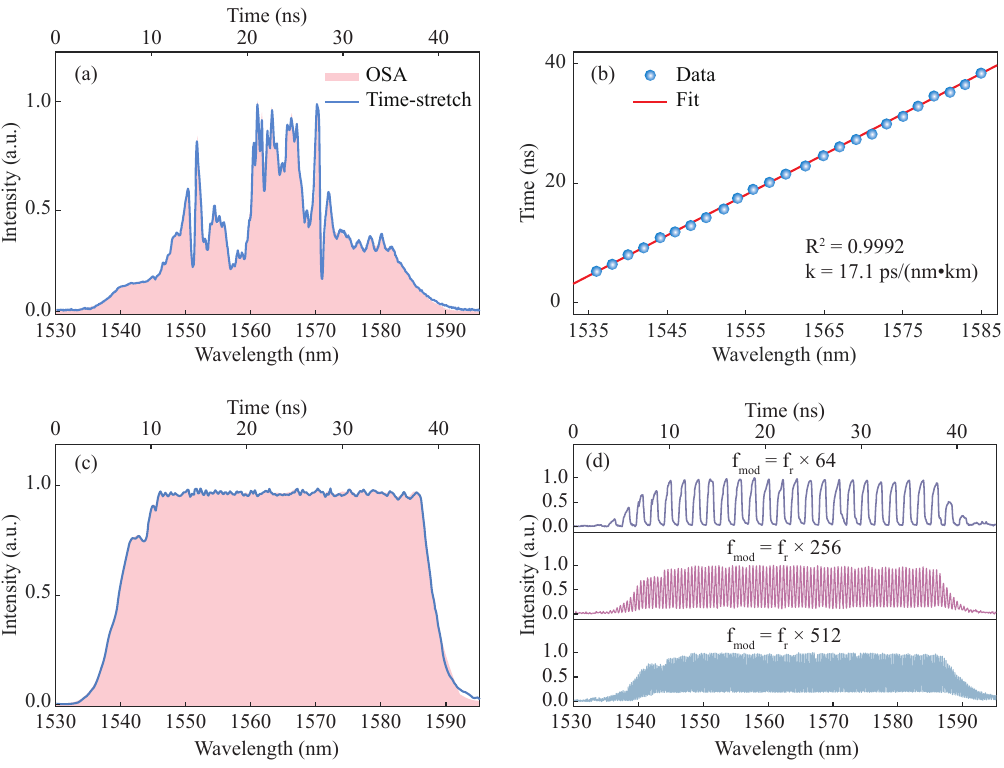}
\caption{Spectro-temporal characterization and calibration. (a) Optical spectrum of the near-infrared signal measured by an optical spectrum analyzer (OSA) with 0.1 nm resolution (shaded area) and the corresponding time-stretched temporal waveform (blue line). (b) Measured time delay as a function of wavelength obtained by scanning a tunable optical filter; the solid line shows a linear fit, confirming a linear wavelength-to-time mapping. (c) Spectrally flattened output of the programmable optical filter (shaded area) and the corresponding temporally stretched waveform (blue line).
(d) Temporally modulated waveforms recorded at different modulation bandwidths of the pulse pattern generator.}
\label{fig3}
\end{figure*}

\begin{figure*}[t!]
\centering
\includegraphics[width=0.85\textwidth]{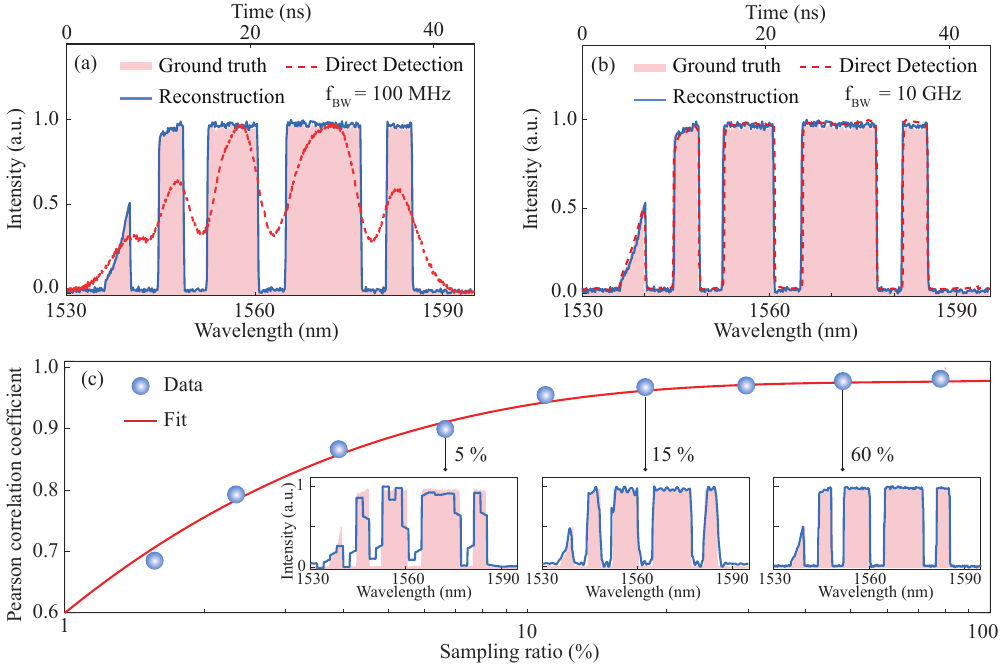}
\caption{Time-stretch computational ghost spectroscopy. (a) Spectrum reconstructed using 512-bit temporal encoding from measurements acquired with a 100 MHz detector (blue solid line). The ground-truth spectrum measured by a 0.1 nm resolution optical spectrum analyzer is shown as the red shaded area, while the red dashed line denotes the direct time-stretch measurement obtained with the same 100 MHz detector. (b) Computationally reconstructed spectrum using 512-bit encoding, compared with the spectrum directly measured by a 10 GHz detector, serving as a high-bandwidth reference. (c) Compressive sensing performance evaluated by the Pearson correlation coefficient as a function of compression ratio. Insets show representative reconstructed spectra at compression ratios of 5\%, 15\%, and 60\%.}
\label{fig4}
\end{figure*}

\begin{figure*}[t!]
\centering
\includegraphics[width=0.8\textwidth]{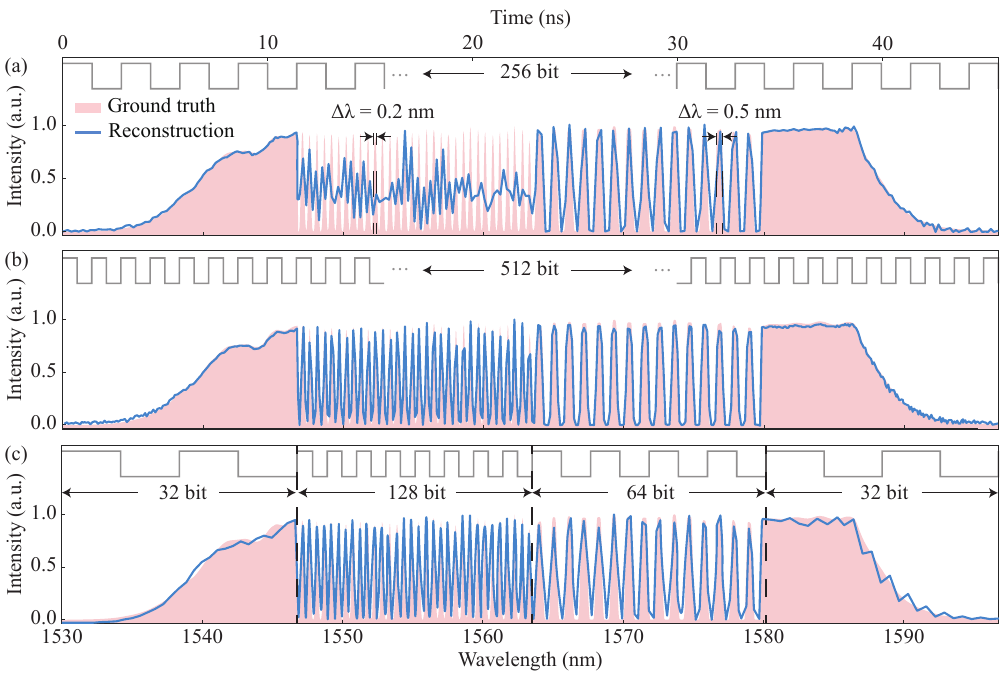}
\caption{High-resolution computational ghost spectroscopy with dynamic sampling rates. (a) Spectrum reconstructed using 256-bit temporal encoding (blue solid line), compared with the ground-truth spectrum measured by a 0.1 nm resolution optical spectrum analyzer (red shaded area). (b) Reconstruction obtained with 512-bit temporal encoding, demonstrating improved spectral resolution. (c) Reconstruction using a warped encoding strategy, in which the wavelength ranges 1530-1546.7 nm, 1546.7-1563.5 nm, 1563.5-1580.5 nm, and 1580.5-1595 nm are encoded with 32, 128, 64, and 32 bits, respectively, enabling enhanced resolution in spectrally dense regions.}
\label{fig5}
\end{figure*}

\section{Results and discussion}

\subsection{Spectro-temporal system calibration}

We first calibrate the spectro-temporal response of the time-stretch system, which underpins all subsequent spectral reconstructions. As shown in Fig. \ref{fig3}(a), a high-fidelity wavelength-to-time mapping is established between the near-infrared spectrum and the temporally stretched waveform. For accurate characterization, the temporal waveform is measured using a 10 GHz InGaAs photodetector, while the corresponding optical spectrum is simultaneously recorded by an optical spectrum analyzer with 0.1 nm resolution.

To quantitatively determine the wavelength-to-time relationship, the programmable optical filter is configured to generate a series of narrowband spectral transmissions with a bandwidth of 2~nm. The measured time delay as a function of the center wavelength is shown in Fig. \ref{fig3}(b), revealing an excellent linear dependence. A linear fit yields a slope of 0.684 ns\,nm$^{-1}$, corresponding to the total GVD of the system.

The native laser spectrum exhibits noticeable intensity non-uniformity, which can compromise detection consistency across spectral channels. To mitigate this effect, the same programmable filter is used to flatten the spectral envelope. The resulting spectrally shaped output and its corresponding time-stretched waveform are shown in Fig. \ref{fig3}(c), confirming a nearly uniform temporal intensity distribution. Figure \ref{fig3}(d) further displays temporally modulated waveforms at different modulation speeds. Although the extinction ratio of the EOM decreases gradually with increasing modulation speed, it remains above 10 dB for all tested conditions.

\subsection{Compressive-sensing ghost spectroscopy}

We next evaluate the performance of the proposed computational spectrometer. In contrast to conventional time-stretch spectroscopy, which relies solely on wavelength-to-time mapping and is therefore constrained by detector bandwidth, our approach combines temporal encoding with computational reconstruction to achieve bandwidth-independent spectral recovery \cite{Kawai2020CommPhys}. In this framework, spectral information is encoded by a set of predefined temporal patterns and retrieved through correlation-based reconstruction, enabling accurate spectroscopy even with low-bandwidth detectors.

The measurement matrix $\mathbf{P}$ is chosen as a Hadamard matrix, whose orthogonality enables efficient multiplexed probing. Since the temporal intensity modulation is inherently non-negative, orthogonal encoding is implemented using a differential acquisition scheme, in which each Hadamard pattern is decomposed into a complementary pair of binary patterns. This strategy not only realizes effective $\pm1$ modulation but also suppresses source intensity fluctuations \cite{Devaux2016Optica}. An additional advantage of the Hadamard basis is that its inverse equals its transpose, allowing fast and numerically stable reconstruction \cite{Zhang2025LPR}. The experimental implementation of these Hadamard patterns via the differential scheme and the resulting temporal modulation fidelity are detailed in Supplementary Note 3, where the applied modulation sequences and the corresponding waveforms of the modulated time-stretched pulses are presented.

Figure \ref{fig4} presents the spectroscopic performance of the system using NIR spectra generated by a programmable optical filter, with the ground truth measured by a 0.1 nm-resolution optical spectrum analyzer. Two measurement modes are compared: direct detection of the time-stretched waveform without modulation, and computational reconstruction using 512-bit temporal encoding. When a 100 MHz detector is employed [Fig. \ref{fig4}(a)], the direct measurement (red dashed curve) suffers from severe distortion and fails to recover meaningful spectral information. In contrast, the computationally reconstructed spectrum (blue solid curve) faithfully reproduces fine spectral features and closely matches the reference spectrum. When the detector bandwidth is increased to 10 GHz [Fig. \ref{fig4}(b)], the direct measurement becomes accurate and serves as a reliable benchmark. The excellent agreement between the computational reconstruction and the high-bandwidth direct measurement confirms both the fidelity of the reconstruction algorithm and the effectiveness of the temporal encoding scheme. These results demonstrate that the proposed spectrometer achieves high-fidelity spectral recovery independent of detector bandwidth, thereby overcoming a fundamental limitation of conventional time-stretch spectroscopy.

We further extend the framework to compressive computational spectroscopy, enabling accurate spectral reconstruction at sub-Nyquist sampling rates \cite{Duarte2008ICASSP}. Instead of requiring a number of measurements equal to the number of spectral channels, the compressive sensing approach exploits the sparsity of typical spectra to substantially reduce acquisition requirements. In the experiment, rows of a Walsh-ordered Hadamard matrix are randomly selected to form the measurement matrix, and spectral reconstruction is performed using the TVAL3 algorithm. Here, the sampling ratio is defined as the ratio between the number of acquired compressive measurements and the total number of spectral channels. The reconstruction quality is quantified by the Pearson correlation coefficient (PCC) with respect to the ground truth. As shown in Fig. \ref{fig4}(c), PCC values exceeding 90\% are achieved even at low sampling ratios, and increase to above 95\% at a 15\% sampling ratio. These results confirm that compressive sensing significantly reduces data acquisition while preserving high reconstruction fidelity, providing a powerful route toward dynamic and high-speed spectroscopic measurements.

\begin{figure*}[t!]
	\centering
	\includegraphics[width=0.8\textwidth]{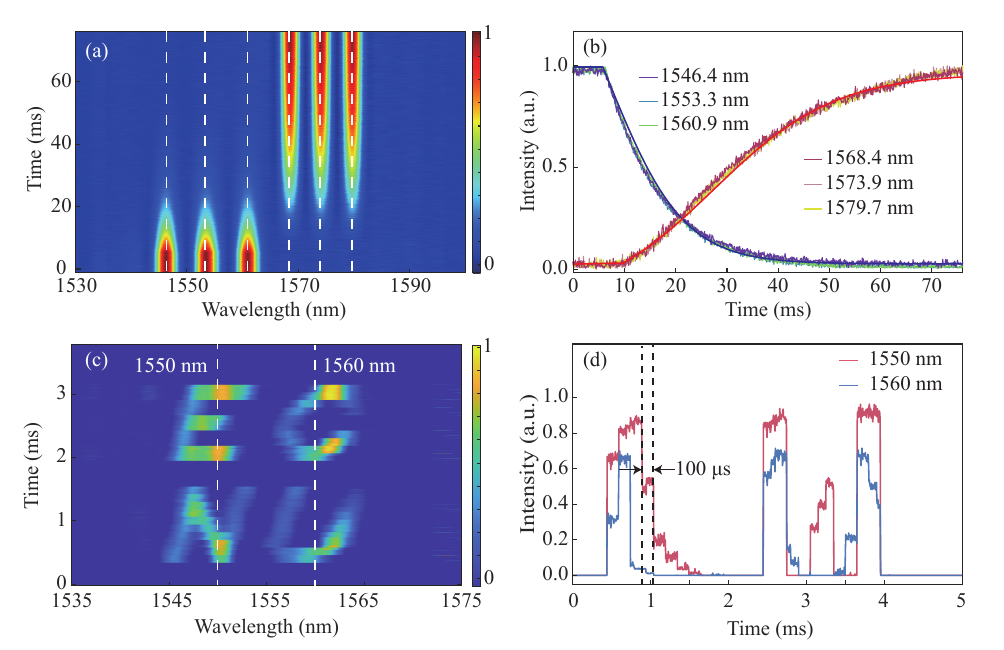}
	\caption{High-speed computational ghost spectroscopy. (a) Time-resolved measurement of a transient spectral switching process, in which a programmable optical filter changes its passbands from 1546.4, 1553.3, and 1560.9 nm (7 nm bandwidth) to 1568.4, 1573.9, and 1579.7 nm (5 nm bandwidth). The measurement employs 256-bit temporal encoding and achieves a frame rate of 42 kHz. (b) Reconstructed output intensity as a function of time at the corresponding wavelengths in (a). (c) Reconstructed spectrum of a digital micromirror device (DMD)-modulated signal using 256-bit encoding with a 20\% compression ratio, achieving a frame rate of 210 kHz. (d) Reconstructed output intensity as a function of time at 1550 nm and 1560 nm for the measurement in (c).}
	\label{fig6}
\end{figure*}

\begin{figure*}[t!]
	\centering
	\includegraphics[width=0.75\textwidth]{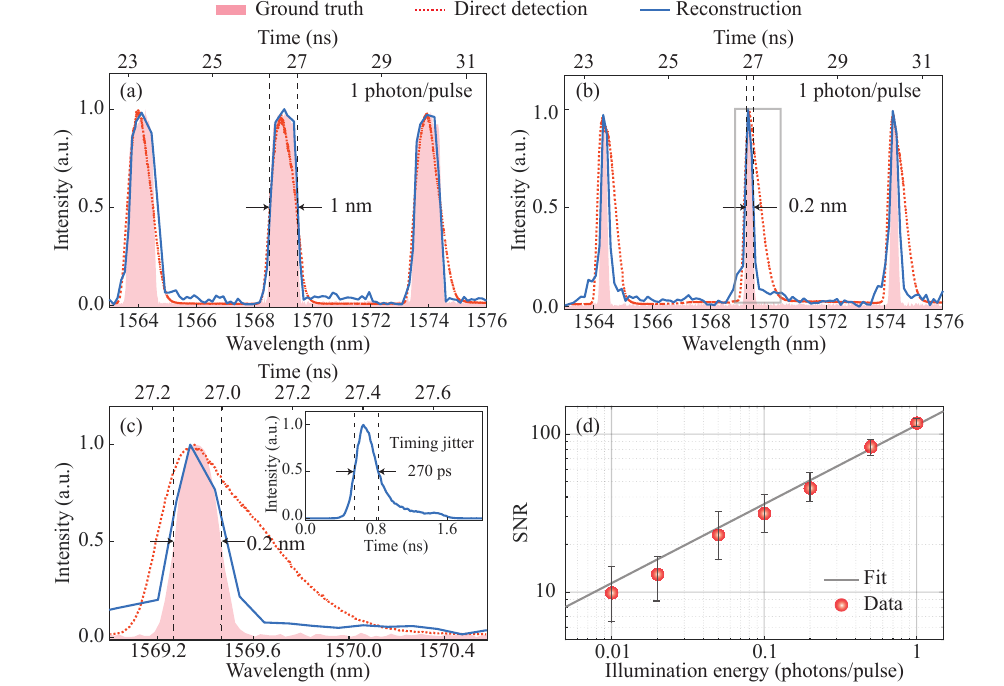}
	\caption{Single-photon computational ghost spectroscopy. (a, b) Spectra reconstructed at the single-photon level using a single-photon avalanche diode (SPAD) with a timing jitter of 270 ps under an illumination flux of 1 photon per pulse. Blue solid lines show the computationally reconstructed spectra at spectral resolutions of 1 nm (a) and 0.2 nm (b), respectively. The acquisition time per encoding pattern is 1 ms. The ground-truth spectra measured by a 0.1 nm resolution optical spectrum analyzer are shown as red shaded areas, while red dashed lines indicate direct time-stretch measurements obtained with the same SPAD. (c) Magnified view of the shaded region in (b); the inset shows the measured timing-jitter distribution of the SPAD. (d) Detection sensitivity of the single-photon computational spectroscopy system, characterized by the signal-to-noise ratio (SNR) as a function of incident photon number per pulse.}
	\label{fig7}
\end{figure*}

\subsection{Computational warped-encoding spectroscopy}

We turn to characterize the spectral resolution of the time-stretch computational spectrometer, which determines its ability to resolve closely spaced spectral features. To quantitatively evaluate the resolution, a programmable optical filter is used to generate narrowband transmission peaks with full widths at half maximum (FWHM) of 0.2 nm and 0.5 nm across the 1530-1590 nm wavelength range. The ground-truth spectra, measured using an optical spectrum analyzer with 0.1 nm resolution, are shown as the red shade areas in Fig. \ref{fig5}, while the corresponding computationally reconstructed spectra are overlaid in blue.

The spectral resolution of the system is fundamentally governed by the temporal resolution of the encoded waveform and the GVD. Specifically, the theoretical resolution is given by $\delta \lambda_{\mathrm{res}} = \delta t_{\mathrm{res}} / (DL)$, where $\delta t_{\mathrm{res}} = T/N_{\mathrm{bit}}$ is the temporal sampling interval, $T$ is the stretched pulse duration, $N_{\mathrm{bit}}$ is the number of encoding bits, $D$ denotes the total GVD, and $L$ denotes the propagation length. The effective temporal resolution is therefore $\delta t_{\mathrm{eff}} = \sqrt{(T/N_{\mathrm{bit}})^2 + \delta t_{\mathrm{jitter}}^2}$, where $\delta t_{\mathrm{jitter}}$ denotes system-level timing jitter. The achievable resolution is fundamentally limited by the modulation bit duration, the GVD, and the system-level timing jitter. Furthermore, the encoding length also affects the data acquisition time, as a longer encoding sequence requires more measurement patterns to complete a full reconstruction.

Figures \ref{fig5}(a) and \ref{fig5}(b) compare reconstruction results obtained using 256-bit and 512-bit temporal encoding, respectively. With 256-bit encoding, the 0.2~nm feature cannot be reliably resolved. In contrast, both the 0.2-nm and 0.5-nm transmission peaks are clearly recovered with 512-bit encoding, demonstrating a substantial enhancement in spectral resolution. These observations are in excellent agreement with the theoretical analysis, highlighting the combined roles of encoding depth and timing stability in determining the system's resolving power.

To further enhance spectral resolution without increasing the total number of measurements, we introduce a warped encoding strategy in the temporal modulation stage. This approach, previously explored in spatial single-pixel imaging \cite{Phillips2017SA}, adaptively redistributes encoding resources by varying the encoding density across temporal regions corresponding to different spectral bands. In warped encoding spectroscopy, encoding resources are allocated according to the local spectral feature complexity, where spectrally dense regions are assigned finer encoding granularity and smoother regions are represented with fewer bits to improve measurement efficiency. In our implementation, the wavelength range is divided into four regions: feature-sparse bands (1530-1546.7 nm and 1580.5-1595 nm) are encoded with 32 bits each, while feature-rich regions (1546.7-1563.5 nm and 1563.5-1580.5 nm) are assigned higher encoding densities of 128 bits and 64 bits, respectively.

This uneven encoding strategy optimizes the utilization of limited encoding resources and effectively enhances spectral resolution where it is most needed. As shown in Fig. \ref{fig5}(c), warped encoding with a total of 256 bits yields a markedly improved reconstruction compared with uniform 256-bit encoding [Fig. \ref{fig5}(a)], while achieving a reconstruction quality comparable to that of uniform 512-bit encoding [Fig. \ref{fig5}(b)]. At the same time, the acquisition time is significantly reduced. It is worth noting that uniform Hadamard encoding does not require prior knowledge of the spectrum and is directly applicable to arbitrary unknown spectral profiles. For completely unknown spectra, an adaptive two-step strategy can be implemented, where a coarse reconstruction is first obtained using a small set of standard Hadamard patterns, followed by refinement of the bit allocation based on this estimate, enabling data-driven optimization without a priori knowledge. These results underscore the adaptability and efficiency of the time-frequency encoded spectroscopy framework, enabling high-resolution and resource-efficient spectral sensing in dynamic or bandwidth-limited scenarios.

\subsection{High-speed computational spectroscopy}

Then, we investigate the high-speed performance of the computational ghost spectrometer. As a first demonstration, we monitor the transient switching dynamics of the liquid-crystal elements in the programmable optical filter. Initially, three narrowband spectral channels centered at 1546.4 nm, 1553.3 nm, and 1560.9 nm with a bandwidth of 7 nm are transmitted. As the driving voltage applied to the liquid-crystal elements is varied, the transmittance of these channels gradually decreases to zero, while three new channels centered at 1568.4 nm, 1573.9 nm, and 1579.7 nm with a bandwidth of 5 nm are simultaneously introduced. Figure \ref{fig6}(a) shows the reconstructed transient spectral evolution obtained using 256-bit temporal encoding. Under this configuration, a frame rate of 42 kHz is achieved, which is primarily determined by the encoding length according to
$\text{frame rate} = f_{\mathrm{r}} / (2N_{\mathrm{bit}})$, where $f_{\mathrm{r}}$ is the laser repetition frequency and $N_{\mathrm{bit}}$ is the number of encoding bits. The temporal transmittance variations at the six representative wavelengths are plotted in Fig. \ref{fig6}(b). The measured rise and fall times of the liquid-crystal switching are 24 ms and 46 ms, respectively, defined by the 10\% - 90\% intensity transition. The excellent agreement among traces at different wavelengths confirms that the switching dynamics of the programmable filter are wavelength-independent. Moreover, the measured transmittance curves are well fitted by hyperbolic tangent functions, indicating the high accuracy and temporal fidelity of the reconstruction.

The acquisition speed can be further increased by incorporating compressive sensing. Using 256-bit encoding with a 20\% compression ratio, the system reaches a frame rate of 210 kHz. To experimentally validate this enhanced capability, we construct a dynamic spectral scene using a digital micromirror device (DMD). In this configuration, a diffraction grating spatially disperses the encoded signal onto the DMD, which performs programmable spectral modulation in the spatial domain. By loading time-varying stripe patterns onto the DMD, a two-dimensional time--wavelength modulation sequence is generated. To more directly visualize the spectral evolution during DMD switching, no spectral flattening using the programmable optical filter is applied in this experiment. As shown in Fig. \ref{fig6}(c), the reconstructed spectral evolution over time faithfully reproduces the letters ``ECNU.'' Because the spectrometer frame rate significantly exceeds the DMD switching speed, transient transitions between successive DMD patterns are clearly resolved. This capability is further confirmed in Fig. \ref{fig6}(d), which shows the temporal intensity traces at 1550 nm and 1560 nm, demonstrating accurate tracking of the 10-kHz DMD switching dynamics.  Furthermore, a lower sampling ratio can be achieved by employing more advanced compressive sensing algorithms or deep learning techniques \cite{Lyu2017SR}, enabling frame rates exceeding 1 MHz. Note that the encoding-based framework requires multiple measurements for each spectral reconstruction, resulting in a reduced effective acquisition rate compared with conventional pulse-by-pulse time-stretch spectroscopy. Nevertheless, this trade-off enables substantially improved robustness and reconstruction fidelity in detector-limited and photon-starved regimes, where sensitivity and relaxed detector requirements become more critical than maximal acquisition speed.

\subsection{High-resolution single-photon spectroscopy}
Finally, we investigate NIR computational spectroscopy operating in the single-photon regime, where detector timing jitter and electronic noise severely limit conventional time-stretch measurements. In single-photon time-stretch spectroscopy, detector timing jitter is directly translated into spectral uncertainty through wavelength-to-time mapping. For InGaAs SPADs, this timing uncertainty originates primarily from the stochastic avalanche buildup process, with commercial devices typically exhibiting timing jitter on the order of 200 ps. Although state-of-the-art devices can approach values near 100 ps, achieving such performance generally requires optimized device structures and more sophisticated fabrication processes \cite{Tosi2014JSTQE, Signorelli2021IEDM}. In our system, the optical signal is precisely attenuated using calibrated neutral-density filters to achieve an average photon flux of approximately one photon per pulse. Under these photon-starved conditions, the analog photodetector is replaced by an InGaAs single-photon avalanche diode (SPAD) with a detection efficiency of 30\% at 1550 nm and a timing jitter of 270 ps. Unlike analog detection, which continuously measures optical intensity, the SPAD generates discrete electrical pulses corresponding to individual photon arrival events. Photon counts are therefore accumulated within predefined integration windows using a photon-counting module, enabling reliable spectral reconstruction at ultralow light levels. Accordingly, the present single-photon experiment is primarily intended to validate the robustness of the proposed framework against timing-jitter-induced spectral degradation, rather than to maximize acquisition speed.

Figures \ref{fig7}(a) and \ref{fig7}(b) compare direct time-stretch measurements and computationally reconstructed spectra for narrowband features with bandwidths of 1 nm and 0.2 nm, respectively. For the 1 nm feature, both approaches closely match the ground-truth spectrum. However, when the spectral feature narrows to 0.2 nm, the direct measurement fails to resolve the true lineshape due to the intrinsic timing jitter of the SPAD. In contrast, the computational reconstruction accurately recovers the narrow spectral feature, as clearly shown in Fig. \ref{fig7}(b). A magnified view of the highlighted region [Fig. \ref{fig7}(c)] further confirms the agreement between the reconstructed spectrum and the ground truth. This result demonstrates an effective enhancement of temporal resolution, in which spectral features beyond the apparent timing resolution of the single-photon detector can be faithfully recovered through correlation-based reconstruction.

To quantitatively assess the detection sensitivity, the signal-to-noise ratio (SNR) of the reconstructed spectra is evaluated as a function of photon flux. The SNR is defined by the absorption peak at 1569.3 nm and the standard deviation of the baseline region without spectral features. Figure \ref{fig7}(d) shows the measured SNR as the average photon flux per pulse is reduced from one photon down to 0.01 photons, with a fixed integration time of 1 s. With $M$ encoding patterns used per reconstructed spectrum, the integration time per pattern is $1/(2M)$. For the data shown in Fig. \ref{fig7}(d), $M$ = 512, giving an integration time of approximately 1~ms per pattern. As expected, the SNR decreases with decreasing photon flux. Remarkably, even at an ultralow level of 0.01 photons per pulse, the reconstructed spectrum maintains an SNR of approximately 10, demonstrating the robustness of the computational approach under extreme photon scarcity. This performance benefits from the multiplexing advantage inherent to the temporal encoding strategy. Unlike conventional point-by-point detection, where each spectral channel collects photons independently, the encoding-based approach exploits global correlations across the modulation sequence. This multiplex gain enhances statistical averaging and stabilizes reconstruction, enabling high-fidelity spectroscopy in the single-photon regime \cite{Sun2024LPR}.

\section{Conclusion}
To date, time-stretch spectroscopy has emerged as a powerful platform for ultrafast spectral measurements, yet its broader application remains fundamentally constrained by the bandwidth and timing jitter of high-speed detection electronics \cite{Goda2013NP, Mahjoubfar2017NP, Kawai2020CommPhys}. These limitations restrict achievable spectral resolution and severely degrade performance in low-light and single-photon regimes \cite{Fard2013LPR, Sun2024LPR}. In this work, we demonstrate a time-stretch computational spectroscopy framework that alleviates these constraints by integrating dispersive wavelength-to-time mapping with programmable temporal encoding and correlation-based reconstruction. By replacing direct temporal waveform sampling with multiplexed encoded acquisition, the proposed approach enables high-fidelity spectral recovery using low-bandwidth and high-jitter detectors. As summarized in Table S1 in Supplementary Note 4, the system achieves a spectral resolution of 0.8 cm$^{-1}$ while significantly relaxing the stringent detector bandwidth and timing-jitter requirements inherent to conventional time-stretch spectroscopy. Furthermore, the multiplexing advantage of Hadamard encoding enables photon-efficient signal integration, allowing reliable spectral reconstruction down to illumination levels of 0.01 photons per pulse in the single-photon regime.

More broadly, as summarized in Table S2 in Supplementary Note 4, the proposed framework should be viewed as offering a different system-level trade-off rather than a universal simplification. Compared with conventional time-stretch spectroscopy \cite{Kawai2020CommPhys, Lei2018NP, Zhou2025Light}, it alleviates stringent detector requirements at the expense of increased demands on temporal modulation, synchronization, and computational reconstruction. Grating-based and FTIR spectrometers provide robust and broadband operation but typically rely on detector arrays or mechanical scanning \cite{Yang2021Science, Zhang2025eLight, Zhu2022JNIS,  Griffiths1983Science, Hashimoto2018NC}, while dual-comb spectroscopy offers exceptional spectral precision and acquisition speed with substantially increased coherence requirements and system complexity \cite{Xu2024Nature, Long2024NP}. Consequently, the present approach is particularly attractive for detector-limited and photon-starved regimes, where relaxed detector requirements and robust spectral recovery are prioritized over maximal acquisition throughput.

The proposed framework is intrinsically extensible and admits several natural performance enhancements. Spectral resolution and bandwidth can be further improved by optimizing the dispersive element and laser repetition rate, enabling finer temporal sampling and higher acquisition speeds \cite{Solli2008NP, Mahjoubfar2017NP}. Increasing the bandwidth of the modulation electrical signal allows for higher-bit-depth encoding, which can further expand the number of detectable spectral channels \cite{Zhang2025LPR}. Moreover, by integrating temporal encoding with single-pixel imaging architectures, the approach can be extended to high-speed hyperspectral imaging, enabling unified spatial-spectral multiplexing for dynamic scene analysis. The associated data burden may be mitigated through compressive sensing and learning-based reconstruction \cite{Meng2024LSA, Sun2026LPR, Park2024PhotoniX, Xu2024NatCommun}. In addition, the technique can also be extended to the mid- and far-infrared regimes via optical frequency upconversion, allowing temporal encoding to be implemented using mature near-infrared modulation technologies \cite{Hashimoto2023LSA, He2024APL, Zhang2025LPR}. Finally, sub-pixel encoding strategies \cite{Hong2023TGRS}, \textit{e.g.}, delay-engineered or interleaved encoding, provide a route to boost the effective spectral resolution.

In summary, we demonstrate a versatile and high-performance NIR spectrometer that achieves a spectral resolution of 0.8 cm$^{-1}$ over a wide spectral range, supports frame rates up to hundreds of kilohertz, and operates efficiently in the single-photon regime. By unifying dispersive mapping, temporal encoding, and algorithmic reconstruction within a flexible and scalable architecture, time-frequency encoded computational spectroscopy establishes a universal platform for high-resolution and photon-efficient spectral measurements. Such capabilities are especially promising for high-sensitivity spectroscopic applications requiring efficient operation under weak-signal and low-photon-flux conditions, including weak-absorption spectroscopy, ultrafast transient spectroscopy, and long-range optical sensing \cite{Hakkel2022NC, Afara2021NP, Jiang2020NP}.

\section*{Acknowledgements}
This work was supported by the Shanghai Pilot Program for Basic Research (TQ20220104); National Natural Science Foundation of China (62505088, 62235019, 62035005); Shanghai Municipal Science and Technology Major Project (2019SHZDZX01); Hainan Provincial Natural Science Foundation of China (126ZD1019, 126QN0836); Postdoctoral Fellowship Program (GZC20250545); China Postdoctoral Science Foundation (2024M760918, 2025T180224); and Fundamental Research Funds for the Central Universities.

\section*{Authors' contributions}
Z.Z., K.H., and H.Z. conceived the idea and designed the experiments. Z.Z. and K.H. conducted the experiments, analyzed the data, and drafted the manuscript. B.S. performed data validation and analysis. B.D. recorded the experimental data. W.Z. assisted with the time synchronization module. Z.Z., K.H., B.S., J.F., and H.Z. revised the manuscript. All authors provided comments and suggestions for improvements.

\section*{Disclosures}
The authors declare no conflict of interest.

\section*{Code and Data Availability}
The data and code that support the findings of this study are available from the corresponding author upon reasonable request.

\section*{Keywords}
Single-photon spectroscopy, temporal ghost imaging, time-stretch measurement, computational spectrometer


\begin{thebibliography}{100}

\bibitem{Ozaki2021Book} Y. Ozaki, C. Huck, S. Tsuchikawa and S. B. Engelsen, Near-Infrared Spectroscopy: Theory, Spectral Analysis, Instrumentation, and Applications, (Springer, Singapore, 2021).

\bibitem{Hakkel2022NC} K. D. Hakkel, M. Petruzzella, F. Ou, A. van Klinken, F. Pagliano, T. Liu, and A. Fiore, ``Integrated near-infrared spectral sensing,'' \textit{Nat. Commun.} \textbf{13}, 103 (2022).

\bibitem{Afara2021NP} I. O. Afara, R. Shaikh, E. Nippolainen, W. Querido, J. Torniainen, J. K. Sarin, and J. T{\"o}yr{\"a}s,  ``Characterization of connective tissues using near-infrared spectroscopy and imaging,'' \textit{Nat. Protoc.} \textbf{16}, 1297-1329 (2021).

\bibitem{Tsuchikawa2022AS} S. Tsuchikawa, T. Ma, and T. Inagaki, ``Application of near-infrared spectroscopy to agriculture and forestry,'' \textit{Anal. Sci.} \textbf{38}, 635--642 (2022).

\bibitem{Yang2021Science} Z. Yang, T. Albrow-Owen, W. Cai, and T. Hasan, ``Miniaturization of optical spectrometers,'' \textit{Science} \textbf{371}, eabe0722 (2021).

\bibitem{Zhang2025eLight} Y. Zhang, E. Yang, H. H. Yoon, Q. Cheng, Z. Sun, T. Hasan, and W. Cai, ``Reconstructive spectrometers: hardware miniaturization and computational reconstruction,'' \textit{eLight} \textbf{5}, 23 (2025).

\bibitem{Zhu2022JNIS} C. Zhu, X. Fu, J. Zhang, K. Qin, and C. Wu, ``Review of portable near infrared spectrometers: current status and new techniques,'' \textit{J. Near Infrared Spectrosc.} \textbf{30}, 51-66 (2022).

\bibitem{Griffiths1983Science} P. R. Griffiths, ``Fourier transform infrared spectrometry,'' \textit{Science} \textbf{222}, 297-302 (1983).

\bibitem{Hashimoto2018NC} K. Hashimoto and T. Ideguchi, ``Phase-controlled Fourier-transform spectroscopy," \textit{Nat. Commun.} \textbf{9}, 4448 (2018).

\bibitem{Xu2024Nature} B. Xu, Z. Chen, T. W. H\"{a}nsch, and N. Picqu\'{e}, ``Near-ultraviolet photon-counting dual-comb spectroscopy,'' \textit{Nature} \textbf{627}, 289--294 (2024).

\bibitem{Long2024NP} D. A. Long, M. J. Cich, C. Mathurin, A. T. Heiniger, G. C. Mathews, A. Frymire, and G. B. Rieker, ``Nanosecond time-resolved dual-comb absorption spectroscopy,'' \textit{Nat. Photonics} \textbf{18}, 127-131 (2024).

\bibitem{Kranendonk2007OptExpress} L. A. Kranendonk, X. An, A. W. Caswell, R. E. Herold, S. T. Sanders, R. Huber, J. G. Fujimoto, Y. Okura, and Y. Urata, ``High speed engine gas thermometry by Fourier-domain mode-locked laser absorption spectroscopy,'' \textit{Opt. Express} \textbf{15}, 15115--15128 (2007).

\bibitem{Huang2022Sensors} D. Huang, Y. Shi, F. Li, and P. K. A. Wai, ``Fourier Domain Mode Locked Laser and Its Applications,'' \textit{Sensors} \textbf{22}, 3145 (2022).

\bibitem{Cai2026arXiv} Z. Cai, Z. Wang, Y. Ding, Y. Wang, C. Wang, C. Yang, Y. Guo, J. Yan, J. Wang, X. Liu, J. Li, R. Zhao, X. Xue, and C. Bao, ``Overcoming sensitivity-bandwidth trade-off in mid-infrared spectroscopy by a microresonator-anchored swept laser,'' \textit{arXiv} arXiv:2602.07757 (2026).

\bibitem{Jiang2020NP} Y. Jiang, S. Karpf, and B. Jalali, ``Time-stretch LiDAR as a spectrally scanned time-of-flight ranging camera,'' \textit{Nat. Photonics} \textbf{14}, 14-18 (2020).

\bibitem{Solli2008NP} D. R. Solli, J. Chou, and B. Jalali, ``Amplified wavelength-time transformation for real-time spectroscopy,'' \textit{Nat. Photonics} \textbf{2}, 48-51 (2008).

\bibitem{Goda2013NP} K. Goda and B. Jalali, ``Dispersive Fourier transformation for fast continuous single-shot measurements,'' \textit{Nat. Photonics} \textbf{7}, 102-112 (2013).

\bibitem{Mahjoubfar2017NP} A. Mahjoubfar, D. V. Churkin, S. Barland, N. Broderick, S. K. Turitsyn, and B. Jalali, ``Time stretch and its applications,'' \textit{Nat. Photonics} \textbf{11}, 341-351 (2017).

\bibitem{Zhang2025NRMP} Y. Zhang, C. Tao, S. Luo, K. Y. Lau, J. Zheng, L. Huang, and Z. Sun, ``Ultra-fast optical time-domain transformation techniques,'' \textit{Nat. Rev. Methods Primers} \textbf{5}, 11 (2025).

\bibitem{Kawai2020CommPhys} A. Kawai, K. Hashimoto, T. Dougakiuchi, V. R. Badarla, T. Imamura, T. Edamura, and T. Ideguchi, ``Time-stretch infrared spectroscopy,'' \textit{Commun. Phys.} \textbf{3}, 152 (2020).

\bibitem{Lei2018NP} C. Lei, H. Kobayashi, Y. Wu, M. Li, A. Isozaki, A. Yasumoto, and K. Goda, ``High-throughput imaging flow cytometry by optofluidic time-stretch microscopy,'' \textit{Nat. Protoc.} \textbf{13}, 1603-1631 (2018).

\bibitem{Zhou2025Light} J. Zhou, L. Mei, M. Yu, X. Ma, D. Hou, Z. Yin, and C. Lei, ``Imaging flow cytometry with a real-time throughput beyond 1,000,000 events per second,'' \textit{Light: Sci. Appl.} \textbf{14}, 76 (2025).

\bibitem{Cai2024SA} Y. Cai, Y. Chen, K. Dorfman, X. Xin, X. Wang, K. Huang, and E. Wu, ``Mid-infrared single-photon upconversion spectroscopy enabled by nonlocal wavelength-to-time mapping,'' \textit{Sci. Adv.} \textbf{10}, eadl3503 (2024).

\bibitem{Yang2022SB} Y. Yang, Y. Jin, X. Xiang, W. Li, T. Liu, S. Zhang, R. Dong, and M. Li, ``Single-photon microwave photonics,'' \textit{Sci. Bull.} \textbf{67}, 700-706 (2022).

\bibitem{Sun2024LPR} B. Sun, K. Huang, H. Ma, J. Fang, T. Zheng, Y. Chu, Y. Liu, Z. Liu, J. Yang, J. Yang, and H. Zeng, ``Single-Photon Time-Stretch Infrared Spectroscopy,'' \textit{Laser Photon. Rev.} \textbf{18}, 2301272 (2024).

\bibitem{Avenhaus2009OL} M. Avenhaus, A. Eckstein, P. J. Mosley, \textit{et al.},
``Fiber-assisted single-photon spectrograph,'' \textit{Opt. Lett.} \textbf{34}, 2873--2875 (2009).

\bibitem{Fard2013LPR} A. M. Fard, S. Gupta, and B. Jalali, ``Photonic time-stretch digitizer and its extension to real-time spectroscopy and imaging,'' \textit{Laser Photonics Rev.} \textbf{7}, 207-263 (2013).

\bibitem{Tiedeck2022ACSPhotonics} S. Tiedeck, M.~B. Heindl, P. Kramlinger, J. Naas, F. Br\"utting, N. Kirkwood, P. Mulvaney, and G. Herink, ``Single-pixel fluorescence spectroscopy using near-field dispersion for single-photon counting and single-shot acquisition,'' \textit{ACS Photonics} \textbf{9}, 2931--2937 (2022).

\bibitem{Hashimoto2023LSA} K. Hashimoto, T. Nakamura, T. Kageyama, V. R. Badarla, H. Shimada, R. Horisaki, and T. Ideguchi, ``Upconversion time-stretch infrared spectroscopy,'' \textit{Light Sci. Appl.} \textbf{12}, 48 (2023).

\bibitem{He2024APL} L. He, H. Wu, W. Wang, B. Hu, X. Yang, and H. Liang, ``Long-wavelength infrared upconversion time-stretch spectroscopy,'' \textit{Appl. Phys. Lett.} \textbf{125}, 071102 (2024).

\bibitem{Redding2013NP} B. Redding, S. F. Liew, R. Sarma, and H. Cao,
``Compact spectrometer based on a disordered photonic chip,'' \textit{Nat. Photonics} \textbf{7}, 746-751 (2013).

\bibitem{Chen2024LSA} C. Chen, H. Gu, and S. Liu, ``Ultra-simplified diffraction-based computational spectrometer,'' \textit{Light Sci. Appl.} \textbf{13}, 9 (2024).

\bibitem{Kong2021NL} L. Kong, Q. Zhao, H. Wang, J. Guo, H. Lu, H. Hao, and P. Wu, ``Single-detector spectrometer using a superconducting nanowire,'' \textit{Nano Lett.} \textbf{21}, 9625-9632 (2021).

\bibitem{Cheng2019NC} R. Cheng, C. L. Zou, X. Guo, S. Wang, X. Han, and H. X. Tang, ``Broadband on-chip single-photon spectrometer,'' \textit{Nat. Commun.} \textbf{10}, 4104 (2019).

\bibitem{Zhang2024NC} Y. Zhang, S. Zhang, H. Wu, J. Wang, G. Lin, and A. P. Zhang, ``Miniature computational spectrometer with a plasmonic nanoparticles-in-cavity microfilter array,'' \textit{Nat. Commun.} \textbf{15}, 3807 (2024).

\bibitem{Xiao2022ACSP} Y. Xiao, S. Wei, J. Xu, R. Ma, X. Liu, X. Zhang, and Z. Wang, ``Superconducting single-photon spectrometer with 3D-printed photonic-crystal filters,'' \textit{ACS Photonics} \textbf{9}, 3450-3456 (2022).

\bibitem{Burghoff2016SciAdv} D. Burghoff, Y. Yang, and Q. Hu, ``Computational multiheterodyne spectroscopy,'' \textit{Sci. Adv.} \textbf{2}, e1601227 (2016).

\bibitem{Ryczkowski2016NP} P. Ryczkowski, M. Barbier, A. T. Friberg, J. M. Dudley, and G. Genty, ``Ghost imaging in the time domain,'' \textit{Nat. Photonics} \textbf{10}, 167--170 (2016).

\bibitem{Devaux2016Optica} F. Devaux, P. A. Moreau, S. Denis, and E. Lantz, ``Computational temporal ghost imaging,'' \textit{Optica} \textbf{3}, 698-701 (2016).

\bibitem{Zhang2025LPR} W. Zhang, K. Huang, X. Wang, B. Sun, J. Fang, Y. Li, and H. Zeng, ``Mid-infrared single-photon computational temporal ghost imaging,'' \textit{Laser Photon. Rev.} \textbf{19}, 2402180 (2025).

\bibitem{Amiot2018OL} C. Amiot, P. Ryczkowski, A. T. Friberg, J. M. Dudley, and G. Genty, ``Supercontinuum spectral-domain ghost imaging,'' \textit{Opt. Lett.} \textbf{43}, 5025-5028 (2018).

\bibitem{Peng2025NC} D. Peng, L. Mei, Z. Gong, Z. Zuo, Z. Liu, Y. Di, S. Liu, J. Wang, C. Guo, X. Liu, H. Song, and G. Yan, ``Single-photon dual-comb ghost imaging spectroscopy,'' \textit{Nat. Commun.} \textbf{16}, 8505 (2025).

\bibitem{Hu2025OL} J. Hu, T. Lv, Z. Wen, W. Huang, M. Yan, and T. B. H. Zeng, ``Broadband coherent Raman spectroscopy based on single-pulse spectral-domain ghost imaging,'' \textit{Opt. Lett.} \textbf{50}, 6201-6204 (2025).

\bibitem{Rabi2021IEEE} S. Rabi, S. Meir, R. Dror, H. Duadi, F. Baldini, F. Chiavaioli, and M. Fridman, ``Spectral ghost imaging for ultrafast spectroscopy,'' \textit{IEEE Photonics J.} \textbf{14}, 1-4 (2021).

\bibitem{Sanna2024AQT} M. Sanna, D. Rizzotti, S. Signorini, and L. Pavesi, ``2 $\mu$m ghost spectroscopy with an integrated silicon quantum photonics source,'' \textit{Adv. Quantum Technol.} \textbf{7}, 2300159 (2024).

\bibitem{Janassek2018PRA} P. Janassek, S. Blumenstein, and W. Els\"{a}\ss er, ``Ghost spectroscopy with classical thermal light emitted by a superluminescent diode,'' \textit{Phys. Rev. Appl.} \textbf{9}, 021001 (2018).

\bibitem{Zhao2024JLT} J. Zhao, Z. Tang, K. Shao, J. Ding, and S. Pan, ``Sub-nanometer resolution spectral-domain computational ghost imaging based on dispersion Fourier transformation,'' \textit{J. Lightwave Technol.} \textbf{43}, 492-498 (2024).


\bibitem{Gibson2020OE} G. M. Gibson, S. D. Johnson, and M. J. Padgett, ``Single-pixel imaging 12 years on: a review,'' \textit{Opt. Express} \textbf{28}, 28190--28208 (2020).

\bibitem{Duarte2008ICASSP} M. F. Duarte, M. A. Davenport, D. Takhar, J. N. Laska, T. Sun, K. F. Kelly and R. G. Baraniuk, ``Single-pixel imaging via compressive sampling,'' \textit{IEEE Signal Process. Mag.} \textbf{25}, 83-91 (2008).


\bibitem{Vaz2020OE} P. G. Vaz, D. Amaral, L. F. R. Ferreira, M. Morgado, J. Cardoso, ``Image quality of compressive single-pixel imaging using different Hadamard orderings," \textit{Opt. Express} \textbf{28}, 11666 (2020).

\bibitem{Bosworth2015OE} B. T. Bosworth, J. R. Stroud, D. N. Tran, T. D. Tran, S. Chin, and M. A. Foster, ``High-speed flow microscopy using compressed sensing with ultrafast laser pulses,'' \textit{Opt. Express} \textbf{23}, 10521--10532 (2015).

\bibitem{Guo2015OE} Q. Guo, H. Chen, Z. Weng, M. Chen, S. Yang, and S. Xie, ``Compressive sensing-based high-speed time-stretch optical microscopy for two-dimensional image acquisition,'' \textit{Opt. Express} \textbf{23}, 29639--29646 (2015).

\bibitem{Mididoddi2017IPJ} C. K. Mididoddi, F. Bai, G. Wang, J. Liu, S. Gibson, and C. Wang, ``High-throughput photonic time-stretch optical coherence tomography with data compression,'' \textit{IEEE Photon. J.} \textbf{9}, 1--15 (2017).

\bibitem{Chi2019IPJ} H. Chi and Z. Zhu, ``Analytical model for photonic compressive sensing with pulse stretch and compression,'' \textit{IEEE Photon. J.} \textbf{11}, 1--10 (2019).

\bibitem{Li2023ACSP} R. Li, Y. Weng, S. Lin, C. Wei, L. Mei, S. Wei, Y. Yao, F. Zhou, D. Wang, K. Goda, and C. Lei, ``All-optical Fourier-domain-compressed time-stretch imaging with low-pass filtering,'' \textit{ACS Photon.} \textbf{10}, 2399--2406 (2023).

\bibitem{Phillips2017SA} D. B. Phillips, M. J. Sun, J. M. Taylor, M. P. Edgar, S. M. Barnett, G. M. Gibson, and M. J. Padgett, ``Adaptive foveated single-pixel imaging with dynamic supersampling,'' \textit{Sci. Adv.} \textbf{3}, e1601782 (2017).

\bibitem{Lyu2017SR} M. Lyu, W. Wang, H. Wang, H. Wang, G. Li, N. Chen, and G. Situ, ``Deep-learning-based ghost imaging,'' \textit{Sci. Rep.} \textbf{7}, 17865 (2017).

\bibitem{Tosi2014JSTQE} A. Tosi, N. Calandri, M. Sanzaro, and F. Acerbi, ``Low-noise, low-jitter, high detection efficiency InGaAs/InP single-photon avalanche diode,'' \textit{IEEE J. Sel. Top. Quantum Electron.} \textbf{20}, 192--197 (2014).

\bibitem{Signorelli2021IEDM} F. Signorelli, F. Telesca, E. Conca, A. Ruggeri, A. Giudice, and A. Tosi, ``InGaAs/InP SPAD detecting single photons at 1550 nm with up to 50\% efficiency and low noise,'' in \textit{Proc. IEEE Int. Electron Devices Meeting (IEDM)}, 20.3.1--20.3.4 (2021).

\bibitem{Meng2024LSA} H. Meng, Y. Gao, X. Wang, X. Li, L. Wang, X. Zhao, and B. Sun, ``Quantum dot-enabled infrared hyperspectral imaging with single-pixel detection,'' \textit{Light Sci. Appl.} \textbf{13}, 121 (2024).

\bibitem{Sun2026LPR} B. Sun, K. Huang, Z. Zhao, B. Dong, J. Fang, and H. Zeng, ``Infrared single-pixel hyperspectral imaging via spatial-temporal multiplexing,'' \textit{Laser Photon. Rev.} \textbf{20}, e01321 (2026).

\bibitem{Park2024PhotoniX} J. Park, and L. Gao, ``Cascaded compressed-sensing single-pixel camera for high-dimensional optical imaging,'' \textit{PhotoniX} \textbf{5}, 37 (2024).

\bibitem{Xu2024NatCommun} Y. Xu, L. Lu, V. Saragadam, and K. F. Kelly, ``A compressive hyperspectral video imaging system using a single-pixel detector,'' \textit{Nat. Commun.} \textbf{15}, 1456 (2024).

\bibitem{Hong2023TGRS} D. Hong, J. Yao, C. Li, D. Meng, N. Yokoya, and J. Chanussot, ``Decoupled-and-coupled networks: self-supervised hyperspectral image super-resolution with subpixel fusion,'' \textit{IEEE Trans. Geosci. Remote Sens.} \textbf{61}, 1-12 (2023).




\end{thebibliography}
\end{document}